\documentclass[twocolumn,showpacs,preprintnumbers,pra]{revtex4}
\usepackage{graphicx}
\usepackage{hyperref}

\begin{document}

\title{Thermal entanglement in one-dimensional Heisenberg quantum spin chains
under magnetic fields}
\author{Shou-Shu Gong and Gang Su$^{\ast}$}
\affiliation{College of Physical Sciences, Graduate University
of Chinese Academy of Sciences, P. O. Box 4588, Beijing 100049,
People's Republic of China}

\begin{abstract}
The thermal pairwise entanglement (TE) of the $S$=$1/2$ XY chain in
a transverse magnetic field is exactly resolved by means of the
Jordan-Wigner transformation in the thermodynamic limit $N$$\rightarrow$$\infty$.
It is found that the TE vanishes at a
fixed point with temperature $T_{c}$$\simeq$$0.4843J$, which is
independent of the magnetic field. A thermal quantity is proposed to witness
the entangled state. Furthermore, the TE of the $S$=$1/2$
antiferromagnetic-ferromagnetic (AF-F) Heisenberg chain is studied
by the transfer-matrix renormalization group method.The TEs of the
spins coupled by AF and F interactions are found to behave
distinctively. The vanishing temperature of the field-induced TE of
the spins coupled by F interactions is observed dependent on the
magnetic field. The results are further confirmed and analyzed
within a mean-field framework.
\end{abstract}

\pacs{03.67.Mn, 03.65.Ud, 75.10.Jm}
\maketitle

Quantum entanglement describes intrinsic correlations incurred in
quantum mechanics. It plays the essential role in quantum
information \cite{NC}, quantum teleportation \cite{CHB}, and quantum
cryptography \cite{AKE}. In condensed matter physics, it provides a
new perspective to understand the collective phenomena in many-body
systems \cite{AFOV}.

The spin entanglement in quantum spin chains is of particular
interest. Many types of entanglement at both zero and finite
temperatures have been extensively studied in various spin systems
(see reviews in Ref. \onlinecite{AFOV}). As the finite-temperature
entanglement (thermal entanglement, TE) can be witnessed
theoretically \cite{BV} and detected experimentally \cite{Ghosh} by
macroscopic variables, most effects have been made to quantify the
TE. The critical temperature (CT) below which the TE survives can
even be estimated in experiment \cite{VB}. Interestingly, the
numerical calculations indicate that the CT of the nearest-neighbor
TE is a fixed point which is independent of the magnetic field in the $S$=$1/2$
Heisenberg chain \cite{ABV} and two-qubit XY spins \cite{W}.
However, the features at the fixed point are unclear and no
explanation exists. It is also a question whether the magnetic field
independence of the CT is a universal phenomenon in quantum spin
chains or there are exceptions. For these questions, in the paper,
the field dependence of the CT of TE in the $S$=$1/2$ XY and AF-F
Heisenberg chains are exactly resolved and studied by means of the
transfer-matrix renormalization group (TMRG) method in the thermodynamic limit $N$$\rightarrow$$\infty$, respectively. An analysis will also be made within the mean-field framework.

The pairwise entanglement of two $S$=$1/2$ spins at sites $i$ and $j$ in the ground state and at finite
temperature can be achieved from the corresponding reduced density matrix $\hat{\rho}_{i,j}$, which,
in the standard basis $\{\mid\uparrow\uparrow\rangle,\mid\uparrow\downarrow\rangle,
\mid\downarrow\uparrow\rangle,\mid\downarrow\downarrow\rangle\}$, can be expressed as
\begin{equation}
\hat{\rho}_{i,j}=\left(
  \begin{array}{cccccccc}
    \langle P^{\uparrow}_{i}P^{\uparrow}_{j}\rangle & \langle P^{\uparrow}_{i}\sigma^{-}_{j}\rangle & \langle \sigma^{-}_{i}P^{\uparrow}_{j}\rangle & \langle \sigma^{-}_{i}\sigma^{-}_{j}\rangle \\
    \langle P^{\uparrow}_{i}\sigma^{+}_{j}\rangle & \langle P^{\uparrow}_{i}P^{\downarrow}_{j}\rangle & \langle \sigma^{-}_{i}\sigma^{+}_{j}\rangle & \langle \sigma^{-}_{i}P^{\downarrow}_{j}\rangle \\
    \langle \sigma^{+}_{i}P^{\uparrow}_{j}\rangle & \langle \sigma^{+}_{i}\sigma^{-}_{j}\rangle & \langle P^{\downarrow}_{i}P^{\uparrow}_{j}\rangle & \langle P^{\downarrow}_{i}\sigma^{-}_{j}\rangle \\
    \langle \sigma^{+}_{i}\sigma^{+}_{j}\rangle & \langle \sigma^{+}_{i}P^{\downarrow}_{j}\rangle & \langle P^{\downarrow}_{i}\sigma^{+}_{j}\rangle & \langle P^{\downarrow}_{i}P^{\downarrow}_{j}\rangle \\
  \end{array}
\right),
\end{equation}
where $P^{\uparrow}$=$\frac{1}{2}(1+\sigma^{z})$,
$P^{\downarrow}$=$\frac{1}{2}(1-\sigma^{z})$, and
$\sigma^{\pm}$=$\frac{1}{2}(\sigma^{x}\pm\sigma^{y})$. The brackets
denote the ground-state and thermodynamic average values at zero and
finite temperatures, respectively, and $\sigma$ are Pauli matrices.
As the phenomenon that is of interest mainly exists in the
nearest-neighbor TE, we shall concentrate only on
$\hat{\rho}_{i,i+1}$ in the following.

The spin operators can be transformed into spinless fermions by the Jordan-Wigner (JW) transformation
\begin{equation}
S^{+}_{i}=c^{\dag}_{i}e^{i\pi\sum_{j<i}c^{\dag}_{j}c_{j}}, \quad
S^{z}_{i}=(c^{\dag}_{i}c_{i}-\frac{1}{2}),
\end{equation}
where $c^{\dagger}_{i}$ and $c_{i}$ are the creation and
annihilation operators of the spinless fermion, respectively.
$\hat{\rho}_{i,i+1}$ becomes
\begin{equation}
\hat{\rho}_{i,i+1}=\left(
  \begin{array}{cccccccc}
  X^{+}_{i} & 0 & 0 & 0 \\
  0 & Y^{+}_{i} & Z^{\ast}_{i} & 0 \\
  0 & Z_{i} & Y^{-}_{i} & 0 \\
  0 & 0 & 0 & X^{-}_{i}
\end{array}
\right), \label{Density}
\end{equation}
where $X^{+}_{i}$=$\langle n_{i}n_{i+1}\rangle$
($n_{i}$$\equiv$$c^{\dag}_{i}c_{i}$), $Y^{+}_{i}$=$\langle
n_{i}(1-n_{i+1})\rangle$, $Y^{-}_{i}$=$\langle n_{i+1}(1-n_{i})\rangle$,
$Z_{i}$=$\langle c^{\dag}_{i}c_{i+1}\rangle$, and $X^{-}_{i}$=$1-\langle
n_{i}\rangle-\langle n_{i+1}\rangle+\langle n_{i}n_{i+1}\rangle$.
As defined, the concurrence of TE of two nearest neighbors is given through
\begin{equation}
\tilde{C}_{i}=\mu_{1}-\mu_{2}-\mu_{3}-\mu_{4}, \label{con}
\end{equation}
\begin{equation}
C_{i}=max\lbrace0,\tilde{C}_{i}\rbrace ,
\end{equation}
where $\mu_{i}$ are the square roots of the eigenvalues of
$\rho_{i,i+1}\tilde{\rho}_{i,i+1}$, where $\mu_{1}$ is the largest.
 $\tilde{\rho}_{i,i+1}$ is a transformed matrix of $\rho_{i,i+1}$, i.e.,
$\tilde{\rho}$=$\sigma_{y}$$\otimes$$\sigma_{y}$$\rho^{\ast}$$\sigma_{y}$$\otimes$$\sigma_{y}$.
Thus, Eq. (\ref{con}) is transformed into
\begin{equation}
\tilde{C}_{i}=2(|Z_{i}|-\sqrt{X^{+}_{i}X^{-}_{i}}). \label{concurrence}
\end{equation}
The concurrence can be calculated from the local density,
hopping term, and site-site correlations of the fermions.

As the observed field independence of the CT of TE are obtained by
either numerical calculations \cite{ABV} or only for two qubits
\cite{W}, a deep understanding is indeed necessary. Therefore, we
shall exactly resolve the concurrence of the $S$=$1/2$ XY chain in a
transverse magnetic field within the thermodynamic limit
to investigate this phenomenon
analytically. The Hamiltonian of the $S$=$1/2$ XY chain is given as
\begin{equation}
H=\sum^{N}_{i=1}\frac{1}{2}J(S^{+}_{i}S^{-}_{i+1}+h.c.)-h\sum^{N}_{i=1}S^{z}_{i},
\label{Hamiltonian}
\end{equation}
where $J($$>$$0)$ is the coupling, and $h$ is the magnetic field. As
the XY chain with F couplings can be obtained by a unitary
transformation, which rotates the odd-site spins by $\pi$ angle
around the $z$-axis, both the AF and F cases give the same results.
Here we take $J$$>$$0$ for simplicity.

\begin{figure}[tbp]
\includegraphics[angle=90,width=0.8\linewidth,clip]{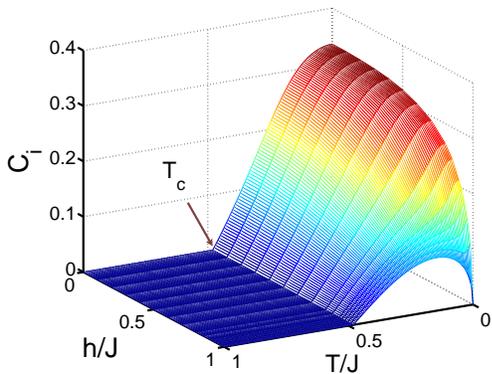}
\caption{(Color online) Temperature and magnetic field dependence of
the thermal entanglement in the $S$=$1/2$ XY chain. The
entanglements vanish at a common temperature $T_{c}$.} \label{XY}
\end{figure}

By applying the JW and Fourier transformations, the Hamiltonian
(\ref{Hamiltonian}) can be diagonalized as
\begin{equation}
H=\sum_{k}(J\cos{k}-h)c^{\dag}_{k}c_{k}=\sum_{k}[\varepsilon(k)-h]c^{\dag}_{k}c_{k},
\label{Hamiltonian2}
\end{equation}
and the elements in the reduced density matrix [Eq. (\ref{Density})] can be explicitly expressed as
\begin{equation}
Z_{i}=\frac{1}{N}\sum_{k}e^{ik}f(k), \quad \langle n_{i}\rangle=\frac{1}{N}\sum_{k}f(k), \label{Z}
\end{equation}
\begin{equation}
X^{+}_{i}=-\frac{1}{N^{2}}\sum_{k_{1},k_{2}}(1-e^{i(k_{1}-k_{2})})\langle
c^{\dag}_{k_{1}}c^{\dag}_{k_{2}}c_{k_{1}}c_{k_{2}}\rangle, \label{X}
\end{equation}
where $f(k)$=$1/(e^{\beta(\varepsilon(k)-h)}+1)$ ($\beta$ is the
inverse temperature and the Boltzmann constant is taken as $k_{B}$=$1$) is the Fermi distribution function.
By the solution of the retarded Green's function $G_{r}(t)$=$\ll c_{k_{1}}(t)c_{k_{2}}(t),
c^{\dag}_{k_{1}}c^{\dag}_{k_{2}}\gg$ ($k_{1}$$\neq$$k_{2}$) \cite{FW},
the expectation value $\langle c^{\dag}_{k_{1}}c^{\dag}_{k_{2}}c_{k_{1}}c_{k_{2}}\rangle$ in
Eq. (\ref{X}) is obtained as
\begin{equation}
\langle
c^{\dag}_{k_{1}}c^{\dag}_{k_{2}}c_{k_{1}}c_{k_{2}}\rangle=-f(k_{1})f(k_{2})
\quad (k_{1}\neq k_{2}),
\end{equation}
and Eq. (\ref{X}) is simplified as
\begin{equation}
X^{+}_{i}=\langle n_{i}\rangle^{2}-Z^{2}_{i}. \label{Xi}
\end{equation}
By substituting Eqs. (\ref{Z}) and (\ref{Xi}) into Eq.
(\ref{concurrence}), $\tilde{C}_{i}$ can be obtained as
\begin{widetext}
\begin{eqnarray}
\tilde{C}_{i}&=&-\frac{2}{\pi}[
\int^{1}_{-1}\frac{xdx}{\sqrt{1-x^{2}}(e^{\beta^{\prime}(x-h^{\prime})}+1)}+
\sqrt{(\int^{1}_{-1}\sqrt{\frac{1+x}{1-x}}\frac{dx}{e^{\beta^{\prime}(x-h^{\prime})}+1})
(\int^{1}_{-1}\sqrt{\frac{1-x}{1+x}}\frac{dx}{e^{\beta^{\prime}(x-h^{\prime})}+1})}
\nonumber \\
&\times&\sqrt{(\frac{1}{\pi}\int^{1}_{-1}\sqrt{\frac{1+x}{1-x}}\frac{dx}{e^{\beta^{\prime}(x-h^{\prime})}+1}-1)
(\frac{1}{\pi}\int^{1}_{-1}\sqrt{\frac{1-x}{1+x}}\frac{dx}{e^{\beta^{\prime}(x-h^{\prime})}+1}-1)}],
\label{C}
\end{eqnarray}
\end{widetext}
where $\beta^{\prime}$=$\beta J$ and $h^{\prime}$=$h/J$. The result
of Eq. ($\ref{C}$) is shown in Fig. \ref{XY}, where the TEs in
different fields vanish at a common CT ($T_{c}$), which is a fixed
point. This common CT indicates that the magnetic field
cannot retrieve the intrinsic TE once it is destroyed by thermal
fluctuations, even though the field changes the TE below the CT.

At $T$=$T_{c}$, $\tilde{C}_{i}$=$0$ and
$Z^{2}_{i}$=$X^{+}_{i}X^{-}_{i}$. Thus, we can derive the equation
\begin{equation}
\langle n_{i}\rangle-\langle n_{i}\rangle^{2}=-\sqrt{2}Z_{i}-Z^{2}_{i} \label{Tc1}
\end{equation}
at $T_{c}$ for any fields. We define $\Phi(\beta,h)$=$\langle
n_{i}\rangle$+$\sqrt{2}Z_{i}$+$Z^{2}_{i}$-$\langle
n_{i}\rangle^{2}$, which can be expressed as
\begin{widetext}
\begin{equation}
\Phi(\beta,h)=\frac{1}{\pi}\int^{1}_{-1}\frac{(1+\sqrt{2}x)dx}{\sqrt{1-x^{2}}(e^{\beta^{\prime}(x-h^{\prime})}+1)}+\frac{1}{\pi^{2}}\int^{1}_{-1}\int^{1}_{-1}
\frac{(xy-1)dxdy}{\sqrt{(1-x^{2})(1-y^{2})}(e^{\beta^{\prime}(x-h^{\prime})}+1)(e^{\beta^{\prime}(y-h^{\prime})}+1)}. \label{Phi}
\end{equation}
\end{widetext}
Thus, $\Phi$$<$$0$ describes the entangled state, and $\Phi$ must be zero at $T_{c}$, i.e., $\Phi(\beta_{c},h)$=$0$, from which $T_{c}$ can be determined. In the absence of magnetic field, $\langle n_{i}\rangle$=$1/2$ at any temperatures, thus, $Z^{T_{c}}_{i,h=0}$=$(1-\sqrt{2})/2$, and $T_{c}$ can be obtained by solving the equation
\begin{equation}
\frac{\sqrt{2}-1}{2}=\frac{J}{\pi T_{c}}\int^{1}_{0}\frac{\sqrt{1-x^{2}}dx}{1+\cosh{\frac{Jx}{T_{c}}}}, \label{integration}
\end{equation}
which indicates that $T_{c}$ is proportional to $J$, i.e., $T_{c}$=$\alpha J$ with $\alpha$$\simeq$$0.4843$.
At $T_{c}$, $\Phi(\beta_{c},h)$ is independent of $h$, i.e., $\partial \Phi(\beta_{c},h)$/$\partial h$=$0$, yielding the equation
\begin{equation}
2(1+\sqrt{2}Z_{i})\frac{\partial{Z_{i}}}{\partial{h}}
=\sqrt{2}(2\langle n_{i}\rangle-1)\frac{\partial{\langle n_{i}\rangle}}{\partial{h}},
\end{equation}
which is satisfied at $T_{c}$ for any fields. This equation, as well
as Eq. (\ref{Tc1}), determine the fixed point completely. In Ref. \onlinecite{W},
the CT for the $S$=$1/2$ two cyclic XY qubits is $0.5673J$, which is larger than the CT
$0.4843J$ in the thermodynamic limit. This is consistent with the result in Ref. \onlinecite{ABV},
where the CT of the spin-$1/2$ Heisenberg chains with $N$=$5$$\sim$$10$ are smaller than that
with $N$=$2$.

In terms of the spin operators, the finite pairwise TE survives when
\begin{equation}
(\langle S^{+}_{i}S^{-}_{i+1}\rangle +\frac{\sqrt{2}}{2})^{2}<\frac{1}{4}+\langle S^{z}_{i}\rangle^{2},
\end{equation}
which indicates that the concurrence of TE is determined by the
competition between the spin fluctuations and local magnetic moment at finite
temperature. As the quantities in Eq. (\ref{Z}) can be expressed as
thermodynamic observables as
\begin{equation}
Z_{i}=\frac{U+Mh}{NJ}+\frac{h}{2J}, \quad  \langle n_{i}\rangle=\frac{M}{N}+\frac{1}{2},
\end{equation}
where $U$=$\langle H\rangle$ is the internal energy, and $M$=$\sum_{i}\langle S_{i}^{z}\rangle$
is the total magnetization, the TE can be witnessed by the negative thermal quantity
\begin{equation}
\Phi(U,M,h)=(\frac{U+Mh}{NJ}+\frac{h}{2J}+\frac{\sqrt{2}}{2})^{2}-(\frac{M}{N})^{2}-\frac{1}{4},
\label{Thermal}
\end{equation}
which can be measured in experiment. Without the field, the magnetization $M$ vanishes, and the TE survives when
\begin{equation}
\frac{\vert U\vert}{NJ}>\frac{\sqrt{2}-1}{2},
\end{equation}
which includes a wider parameter region than the sufficient
condition $\vert U\vert /NJ$$>$$1/4$ for the entangled state that is
proposed for the spin chains with Heisenberg or XY interactions
\cite{BV}. In a magnetic field, Eq. (\ref{Thermal}) also implies
that the witness $\vert U+Mh\vert /NJ$$>$$1/4$ proposed in Ref.
\onlinecite{BV} can be improved to cover wider parameter region for
the entangled state. The exact solution not only confirms that the
CT of the intrinsic TE which survives in the absence of field is a fixed
point, but also reveals some features at the fixed point
from the perspectives of the local spin competition and macroscopic
thermodynamic behavior.

To investigate the possible exceptions of the field independence of
the CT, we study the TE in an $S$=$1/2$ AF-F alternating Heisenberg
chain by means of the TMRG. The Hamiltonian of this alternating
chain is given by
\begin{equation}
H=\sum\limits_{j}(J_{\mathrm{a}}\mathbf{S}_{2j-1}\cdot
\mathbf{S}_{2j}+J_{\mathrm{f}}\mathbf{S}_{2j}\cdot
\mathbf{S}_{2j+1})-h\sum\limits_{j}S_{j}^{z}, \label{Ham}
\end{equation}
where $J_{\mathrm{a}}$$>$$0$, $J_{\mathrm{f}}$$<$$0$
denote the AF and F couplings, respectively. $J_{\mathrm{a}}$ is taken as the
energy scale and $J_{\mathrm{f}}$/$J_{\mathrm{a}}$=$-1$. This AF-F chain has a Haldane gap
$\Delta$$\simeq$$0.6J_{a}$ in the ground state \cite{Hida}, and the
saturation field $h_{s}$$\simeq$$1.1J_{a}$. In experiment, this
model has been realized and studied extensively \cite{EXAFF}. The
TMRG \cite{TMRG} method is a powerful tool for studying the
thermodynamics of one-dimensional quantum systems in the thermodynamic limit \cite{Group}. In
our calculations, the width of the imaginary time slice is taken as
$\varepsilon$=$0.1$, and the error caused by the Trotter-Suzuki
decomposition is less than $10^{-3}$. During the TMRG iterations,
$60$ states are retained, and the truncation error is less than
$10^{-6}$.

\begin{figure}[tbp]
\includegraphics[width=1.0\linewidth,clip]{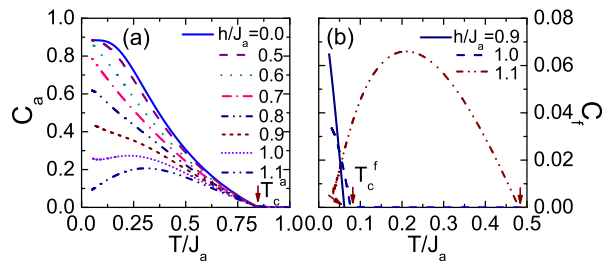}
\caption{(Color online) Temperature dependence of the thermal
entanglement of (a) $C_{a}$, and (b) $C_{f}$ for the $S$=$1/2$ AF-F chain at various fields
obtained by means of the TMRG.} \label{AFF}
\end{figure}

Owing to the alternation of the couplings, the TEs of the spins
coupled by $J_{\mathrm{a}}$ and $J_{\mathrm{f}}$, which are denoted
as $C_{a}$ and $C_{f}$, respectively, are expected to be distinct.
Figure \ref{AFF}(a) shows the temperature dependence of the TE
$C_{a}$ at different fields. It is shown that in the absence of the
field, the intrinsic TE survives, and $C_{a}$ vanishes at
$T^{a}_{c}$$\simeq$$0.85J_{a}$ due to thermal fluctuations. In the
presence of the field, the CT keeps invariant. Although the
alternation is involved, the CT of the intrinsic TE is still a fixed point.

However, the entanglement induced by the field does not comply such
a rule. For the F couplings, the TE of the spins coupled by $J_{f}$
is absent without a field. When the applied field closes the gap and
increases up to about $0.9J_{a}$, the TE is induced by the field, as
shown in Fig. \ref{AFF}(b). With further increasing the field, the
CT of the field-induced TE enhances to reach the maximum at the
saturation field. A further increase of the field makes the
field-induced TE vanish as the spins are fully polarized at zero
temperature. It can be seen that, different from the intrinsic TE,
the CT of the field-induced TE is dependent on the magnetic field.

Furthermore, the intrinsic TE of the $S$=$1/2$ AF-AF-AF-F tetrameric Heisenberg
chain \cite{GSS} is studied using the TMRG, which are not presented
here. It is shown that the CT retains a fixed point,
which is also observed in the trimerized F-F-AF chain \cite{Sun}.
The observations suggest that the CT of the intrinsic TE in one-dimensional
Heisenberg antiferromagnets might be a fixed point.

\begin{figure}[tbp]
\includegraphics[width=1.0\linewidth,clip]{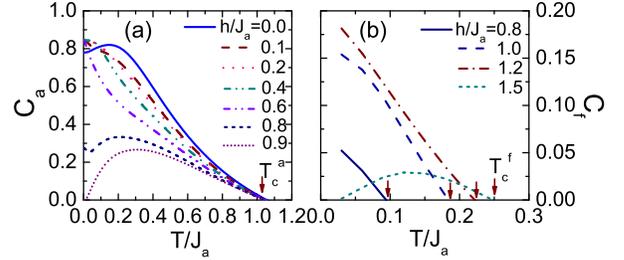}
\caption{(Color online) Temperature dependence of the thermal
entanglement of (a) $C_{a}$, and (b) $C_{f}$ for the $S$=$1/2$ AF-F
chain at various fields obtained by the mean-field theory.}
\label{AFFMean}
\end{figure}

Next, we treat the AF-F chain within the mean-field framework, which
may extend the discussions to the general $S$=$1/2$ alternating
Heisenberg antiferromagnetic chains with nearest-neighbor interactions.
Following the steps in Ref. \onlinecite{YF}, we make the Hartree-Fock
approximation to the Hamiltonian (\ref{Ham}) after the JW and
Fourier transforms, and obtain the mean-field Hamiltonian after
omitting a constant:
\begin{eqnarray}
H_{HF}&=&\sum_{k}\lbrace[(J_{\mathrm{a}}+J_{\mathrm{f}})(d_{b}-\frac{1}{2})-h]a^{\dagger}_{k}a_{k}
\nonumber \\
&+&[(J_{\mathrm{a}}+J_{\mathrm{f}})(d_{a}-\frac{1}{2})-h]b^{\dagger}_{k}b_{k}\rbrace \nonumber \\
&+&\sum_{k}[J_{\mathrm{a}}(\frac{1}{2}-p_{ab})e^{ik/2}a^{\dagger}_{k}b_{k}+h.c. \nonumber \\
&+&J_{\mathrm{f}}(\frac{1}{2}-p_{ba})e^{ik/2}b^{\dagger}_{k}a_{k}+h.c.],
\end{eqnarray}
where $d_{a}$=$\langle a^{\dagger}_{j}a_{j} \rangle$,
$d_{b}$=$\langle b^{\dagger}_{j}b_{j} \rangle$, $p_{ab}$=$\langle
b^{\dagger}_{j}a_{j} \rangle$, $p_{ba}$=$\langle
a^{\dagger}_{j+1}b_{j} \rangle$, which are obtained by
self-consistent calculations. Then the Bogoliubov transformation is
taken to diagonalize the above Hamiltonian. Thus, the TE can be calculated
from the quasiparticle representation. Figure \ref{AFFMean} shows
the mean-field results of the TE $C_{a}$ and $C_{f}$. It is shown
that although the values of the critical fields and CT are not
accurate, the mean-field results still preserve the features of
the CT. The intrinsic TE $C_{a}$ vanishes at a common CT, while the
field-induced TE $C_{f}$ is dependent on the field. As shown in Fig.
\ref{AFFMean}(b), the CT of $C_{f}$ enhances with increasing the
field until to the maximum at the saturation field, which is analogous to the TMRG
result.

In this fermion mapping, $Z_{ab}$=$p_{ab}^{\ast}$, and
$X^{+}_{ab}$=$d_{a}d_{b}-Z_{ab}^{2}$, where $Z_{ab}$ and
$X^{+}_{ab}$ are the values defined in Eq. (\ref{Density}) of the
spins coupled by $J_{a}$. Thus, the concurrence $C_{a}$ can be
expressed by Eq. (\ref{concurrence}) using these quantities. At the
CT ($T_{c}^{a}$), we have
\begin{equation}
[\vert p_{ab}\vert^{2}-d_{a}(d_{b}-1)][\vert p_{ab}\vert^{2}-d_{a}(d_{b}-1)]=2\vert p_{ab}\vert^{2}.
\end{equation}
The calculations show that $p_{ab}$ is real, and $d_{a}$=$d_{b}$. Thus, the above
equation can be simplified as
\begin{equation}
d_{a}-d_{a}^{2}=-\sqrt{2}p_{ab}-p_{ab}^{2},
\end{equation}
which has the same form as that of the XY chain [Eq. (\ref{Tc1})],
yielding the following inequality
\begin{equation}
(\langle
S^{+}_{2j-1}S^{-}_{2j}\rangle+\frac{\sqrt{2}}{2})^{2}<\frac{1}{4}+\langle
S^{z}_{2j-1}\rangle^{2}
\end{equation}
for the entangled $C_{a}$. For the field-induced TE $C_{f}$, we have
$d_{a}-d_{a}^{2}$=$-\sqrt{2}p_{ba}-p_{ba}^{2}$ at the CT, which
is satisfied at different CTs for different fields. The
entangled $C_{f}$ is described by $(\langle
S^{+}_{2j}S^{-}_{2j+1}\rangle$+$\frac{\sqrt{2}}{2})^{2}$$<$$\frac{1}{4}$+$\langle
S^{z}_{2j}\rangle^{2}$. Note that the thermal quantity witness cannot
be written in a form as simple as Eq. (\ref{Thermal}) within the present
self-consistent calculations.

In summary, we have studied the field dependence of the CT of TE in
the $S$=$1/2$ spin chains within the thermodynamic limit $N$$\rightarrow$$\infty$.
The concurrence of the TE in the
spin-$1/2$ XY chain is exactly resolved. It is found that the CT of the
TE is a fixed point. An equation is given to
determine the CT, which is found to be $T_{c}$$\simeq$$0.4843J$ and smaller
than that of the two-qubit system. The
thermal witness for the entangled state is also proposed.
Furthermore, the TE of an $S$=$1/2$ AF-F chain is studied by means
of the TMRG method and mean-field treatment, which indicates that
the CT of the intrinsic TE of the spins coupled by AF couplings is a
fixed point, while that of the field-induced TE of the
spins coupled by F couplings changes with the field. The exact
solution of the XY chain as well as the mean-field result of the
AF-F chain indicate that the disappearance of the TE is determined by
the competition between the spin fluctuations and local magnetic
moment at finite temperatures. The observations suggest that
it may be a general phenomenon in one-dimensional Heisenberg antiferromagnets
that the CT of the intrinsic TE is a fixed point independent of the magnetic field.

\acknowledgments

This work is supported in part by the National Science Fund for
Distinguished Young Scholars of China (Grant No. 10625419), the
National Science Foundation of China (Grants No. 90403036 and No.
20490210), the MOST of China (Grant No. 2006CB601102), and the
Chinese Academy of Sciences.

\end{document}